\documentclass[aps,pra,twocolumn,amsmath,amssymb,showpacs]{revtex4}

\newcommand{\bra}[1]{\langle#1|}
\newcommand{\ket}[1]{|#1\rangle}
\usepackage[dvips]{graphicx}
\usepackage{mathrsfs}

\begin{document}

\title{Error models for mode-mismatch in linear optics quantum computing}

\author{Peter P. Rohde}
\email[]{rohde@physics.uq.edu.au}
\homepage{http://www.physics.uq.edu.au/people/rohde/}
\author{Timothy C. Ralph}
\affiliation{Centre for Quantum Computer Technology, Department of Physics\\ University of Queensland, Brisbane, QLD 4072, Australia}

\date{\today}

\begin{abstract}
One of the most significant challenges facing the development of linear optics quantum computing (LOQC) is mode-mismatch, whereby photon distinguishability is introduced within circuits, undermining quantum interference effects. We examine the effects of mode-mismatch on the parity (or fusion) gate, the fundamental building block in several recent LOQC schemes. We derive simple error models for the effects of mode-mismatch on its operation, and relate these error models to current fault tolerant threshold estimates.
\end{abstract}

\pacs{03.67.Lx,42.50.-p}

\maketitle

\section{Introduction}
Linear optics quantum computing (LOQC), as it was originally proposed \cite{bib:KLM01}, suffered the problem of unfavorable scaling in physical resource requirements. Recently, several proposals have been made which significantly reduce these requirements. Most notably, schemes employing cluster states \cite{bib:Raussendorf01,bib:Raussendorf03,bib:Nielsen04,bib:BrowneRudolph05,bib:Varnava05} and parity states \cite{bib:GilchristHayes05,bib:RalphHayes05} have been suggested. The fundamental building block of many such schemes is the parity gate \cite{bib:Weinfurter94,bib:BraunsteinMann95} (also referred to as the type-II fusion gate \cite{bib:BrowneRudolph05}), which projects an incident two-photon state into the even- or odd-parity sub-space.

\begin{figure}[!htb]
\includegraphics[width=0.6\columnwidth]{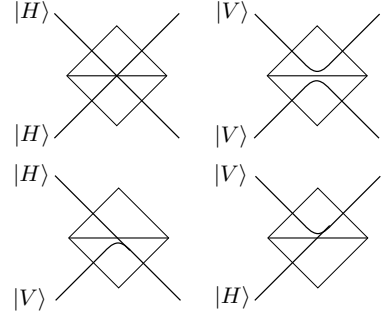}
\caption{Parity measurement using a PBS, which completely transmits horizontally- and completely reflects vertically polarized photons. Upon post-selection on detecting exactly one photon at each beamsplitter output, only the $\ket{HH}$ and $\ket{VV}$ components survive, where $H$ and $V$ denote horizontal- and vertical polarization. Detection is performed in the diagonal/anti-diagonal basis. This effectively destroys all $H/V$ information, preventing the detectors from distinguishing between the $\ket{HH}$ and $\ket{VV}$ cases. Thus, the procedure coherently projects incident two-photon states into the sub-space spanned by $\ket{HH}$ and $\ket{VV}$ -- the even-parity sub-space. This occurs non-deterministically, with success probability depending on the magnitude of the even-parity terms in the incident state. A projection into the odd-parity sub-space can be implemented trivially by first performing a bit-flip operation on one of the incident photons.} \label{fig:PBS_measurement}
\end{figure}

The parity gate is implemented physically using a polarizing beamsplitter (PBS) and post-selection, described in Fig.~\ref{fig:PBS_measurement}. Parity measurement has many applications and, for example, forms the basis of the linear optics controlled-{\sc NOT} ({\sc CNOT}) gate described in Ref.~\cite{bib:Pittman01}, the entanglement purification protocol of Ref.~\cite{bib:Pan01}, the cluster state LOQC scheme of Ref.~\cite{bib:BrowneRudolph05} and the parity encoded LOQC scheme of Refs.~\cite{bib:RalphHayes05,bib:GilchristHayes05}.

One of the most significant challenges facing the experimental realization of LOQC circuits is mode-mismatch \cite{bib:RohdeRalph05,bib:RohdePryde05,bib:RohdeRalph05b}, whereby photon indistinguishability is compromised within a circuit, undermining the desired quantum interference effects. In this paper we consider the effects of mode-mismatch on the parity gate and derive a general error model describing these effects. We apply this model specifically to the the cluster state approach to LOQC. Our results suggest that in this context mode-mismatch can be tolerated using existing fault tolerance techniques for dealing with general depolarizing noise. We relate physical parameters, such as the degree of mode-mismatch and photo-detector characteristics, to current fault tolerance threshold estimates.

\section{General error model}
Consider an arbitrary $n$-qubit, polarization encoded state. This can be expressed generally in the form
\begin{eqnarray}
\ket{\psi}=\sum_{i_1,\dots,i_n\in\{H,V\}}\lambda_{i_1,\dots,i_n}\ket{i_1}\dots\ket{i_n}
\end{eqnarray}
Expanding around the first two qubits, an equivalent expression is
\begin{eqnarray} \label{eq:state_rep}
\ket{\psi}&=&\alpha_{HH}\ket{HH}\ket{\phi_{HH}}+\alpha_{HV}\ket{HV}\ket{\phi_{HV}}\nonumber\\
&+&\alpha_{VH}\ket{VH}\ket{\phi_{VH}}+\alpha_{VV}\ket{VV}\ket{\phi_{VV}}
\end{eqnarray}
where the $\alpha_{xy}$ coefficients denote the amplitude of the corresponding terms and $\ket{\phi_{xy}}$ the state of the rest of the system for the respective state of the first two qubits.

We wish to perform the parity gate between the two factored qubits. We express these qubits in terms of their temporal wave-functions, $\psi_A(t)$ and $\psi_B(t)$,
\begin{widetext}
\begin{eqnarray}
\ket{\psi}&=&\alpha_{HH}\left(\int_{-\infty}^{\infty}\psi_A(t)\hat{a}^\dag_H(t)\,\mathrm{d}t\ket{0}\right)\left(\int_{-\infty}^{\infty}\psi_B(t)\hat{b}^\dag_H(t)\,\mathrm{d}t\ket{0}\right)\ket{\phi_{HH}}\nonumber\\
&+&\alpha_{HV}\left(\int_{-\infty}^{\infty}\psi_A(t)\hat{a}^\dag_H(t)\,\mathrm{d}t\ket{0}\right)\left(\int_{-\infty}^{\infty}\psi_B(t)\hat{b}^\dag_V(t)\,\mathrm{d}t\ket{0}\right)\ket{\phi_{HV}}\nonumber\\
&+&\alpha_{VH}\left(\int_{-\infty}^{\infty}\psi_A(t)\hat{a}^\dag_V(t)\,\mathrm{d}t\ket{0}\right)\left(\int_{-\infty}^{\infty}\psi_B(t)\hat{b}^\dag_H(t)\,\mathrm{d}t\ket{0}\right)\ket{\phi_{VH}}\nonumber\\
&+&\alpha_{VV}\left(\int_{-\infty}^{\infty}\psi_A(t)\hat{a}^\dag_V(t)\,\mathrm{d}t\ket{0}\right)\left(\int_{-\infty}^{\infty}\psi_B(t)\hat{b}^\dag_V(t)\,\mathrm{d}t\ket{0}\right)\ket{\phi_{VV}}
\end{eqnarray}
\end{widetext}
where $\hat{a}^\dag(t)$ and $\hat{b}^\dag(t)$ are the time-specific creation operators for the first two photons. Note that while we specifically make reference to  temporal wave-functions, the same arguments hold in any photonic degree of freedom, such as spatial or spectral.

Next we apply the parity gate between qubits $A$ and $B$ and post-select upon detecting exactly one photon at each beamsplitter output, the required \emph{success} signature. Measurements are modeled using the photo-detector model described in Ref.~\cite{bib:RohdeRalph05c}. In this model, photo-detectors are characterized by two parameters -- their \emph{resolution} ($\delta$) and \emph{bandwidth} ($\Delta$). The resolution characterizes the spectral uncertainty in a measurement event and the detector is unable to distinguish between spectral components within this range. The bandwidth characterizes the total range of frequencies the detector responds to. See Ref.~\cite{bib:RohdeRalph05c} for a complete description and physical motivation. Based on this model, each measurement can be expressed generally in the form
\begin{widetext}
\begin{equation}
\hat\rho_\mathrm{measured}=\mathrm{tr}_D\left[\int_{-\Delta}^{\Delta}\left(\int_{\omega_0-\delta}^{\omega_0+\delta}\ket{\omega}_D\bra{\omega}_D\,\mathrm{d}\omega\right)\hat\rho_\mathrm{in}\left(\int_{\omega_0-\delta}^{\omega_0+\delta}\ket{\omega}_D\bra{\omega}_D\,\mathrm{d}\omega\right)\mathrm{d}\omega_0\right]\nonumber\\
\end{equation}
\end{widetext}
where $\ket\omega_D\bra\omega_D$ is the projector onto the frequency eigenstate $\omega$, acting on photon $D$, the one being detected. $\hat\rho_\mathrm{in}$ is the incident state and $\hat\rho_\mathrm{measured}$ is the state following photo-detection.

\section{Error model for the parity gate}
In the case of parity measurement the output state can be expressed in the form
\begin{eqnarray}
f_{II}\ket{\psi}&=&|\alpha_{HH}|^2\ket{\phi_{HH}}\bra{\phi_{HH}}\nonumber\\
&+&\gamma\alpha_{HH}\alpha_{VV}^*\ket{\phi_{HH}}\bra{\phi_{VV}}\nonumber\\
&+&\gamma\alpha_{VV}\alpha_{HH}^*\ket{\phi_{VV}}\bra{\phi_{HH}}\nonumber\\
&+&|\alpha_{VV}|^2\ket{\phi_{VV}}\bra{\phi_{VV}}
\end{eqnarray}
where $f_{II}$ denotes the parity gate operation and normalization factors have been omitted for simplicity. $\gamma$ is a coherence parameter, and is non-trivially related to the detectors' resolution and bandwidth, as well as the integral overlap of the interacting photons' wave-functions which characterizes the degree of mode-mismatch between them. $\gamma$ obeys $0\leq\gamma\leq1$, where $\gamma=1$ corresponds to complete photon indistinguishability, the ideal case, and $\gamma=0$ to complete photon distinguishability.

In the case of ideal photo-detectors (\emph{i.e.} infinite bandwidth and zero resolution, see Ref.~\cite{bib:RohdeRalph05c}), $\gamma$ is equal to the integral overlap of the interacting photons,
\begin{equation}
\gamma_\mathrm{ideal}=\left|\int_{-\infty}^\infty\int_{-\infty}^\infty\psi_A(\omega_A)\psi_B^*(\omega_B)\mathrm{d}\omega_A\mathrm{d}\omega_B\right|^2
\end{equation}
and is related to the Hong-Ou-Mandel (HOM) \cite{bib:HOM87} visibility \footnote{HOM interference is a physically different scenario, where two photons interact on a 50/50 beamsplitter rather than a PBS. We include this relationship because HOM visibility is a commonly quoted measure of mode-mismatch and therefore gives some insight into what values for $\gamma$ are realistically achievable.} by
\begin{equation}
\gamma_\mathrm{ideal}=\frac{2V}{1+V}
\end{equation}

\begin{figure}[!htb]
\includegraphics[width=0.7\columnwidth]{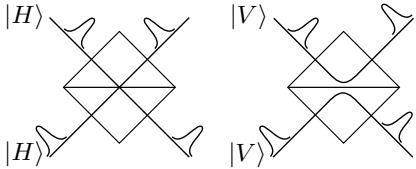}
\caption{Parity measurement in the presence of mode-mismatch (graphically represented as a temporal displacement in one of the incident photons). Now the detection process reveals \emph{some} information as to whether an incident state was $\ket{HH}$ or $\ket{VV}$. For example, in the illustration shown, if the upper photo-detector \emph{clicks} after the lower one, the \emph{which-path} information allows us to retrodict that the incident state was $\ket{HH}$ with greater likelihood than $\ket{VV}$. While the process still projects incident states into the even-parity sub-space, it no longer does this coherently due to the presence of this classical information. The projection performed in the \emph{failure} cases is unaffected by mode-mismatch since complete \emph{which-path} information already exists.} \label{fig:PBS_measurement_MM}
\end{figure}

Ideally, the output state is the coherent superposition
\begin{equation} \label{eq:psi_ideal}
\ket{\psi_\mathrm{ideal}}=\alpha_{HH}\ket{\phi_{HH}}+\alpha_{VV}\ket{\phi_{VV}}
\end{equation}
In the presence of mode-mismatch, $\gamma\leq1$, the output state decoheres into a mixture of this state and the corresponding negative superposition
\begin{equation} \label{eq:psi_error}
\ket{\psi_\mathrm{error}}=\alpha_{HH}\ket{\phi_{HH}}-\alpha_{VV}\ket{\phi_{VV}}\end{equation}
where the degree of decoherence is realated to $\gamma$. The output state can be expressed in the form
\begin{equation} \label{eq:parity_error_model}
f_{II}\ket{\psi}=(1-p_\mathrm{error})\ket{\psi_\mathrm{ideal}}\bra{\psi_\mathrm{ideal}}+p_\mathrm{error}\ket{\psi_\mathrm{error}}\bra{\psi_\mathrm{error}}
\end{equation}
where $p_\mathrm{error}$ is the error probability. This error model can be understood intuitively according to Fig.~\ref{fig:PBS_measurement_MM}.
$p_\mathrm{error}$ can be expressed in terms of $\gamma$,
\begin{equation}
p_\mathrm{error}=\frac{1-\gamma}{2}
\end{equation}

\begin{figure*}[!htb]
\includegraphics[width=\textwidth]{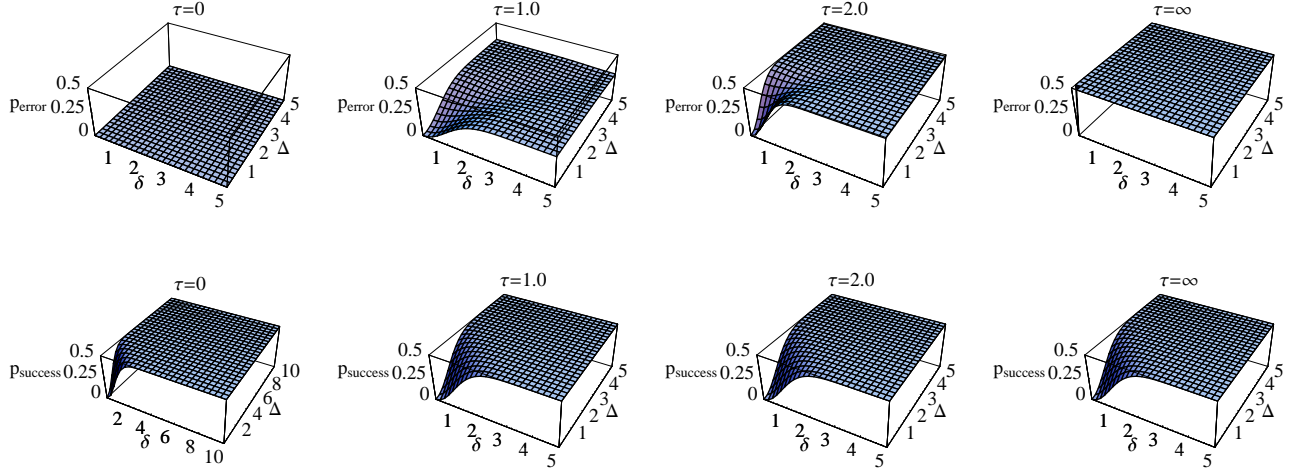}
\caption{Error probability ($p_\mathrm{error}$) and probability of detecting the \emph{success} signature ($p_\mathrm{success}$) against the photo-detectors' spectral resolution ($\delta$) and bandwidth ($\Delta$) for various degrees of temporal mode-mismatch ($\tau$). All quantities are in units of photon bandwidth (\emph{i.e.} variance).} \label{fig:p_delta}
\end{figure*}

The relationship between the degree of mode-mismatch, detector characteristics and error probability is shown in Fig.~\ref{fig:p_delta}. We assume photons have transform-limited Gaussian temporal wave-packets. Fig.~\ref{fig:p_delta} indicates that if error rates are to be minimized, photo-detector bandwidth and resolution ought to be kept as small as possible. This can be achieved through narrowband filtering, a technique currently employed in many coincidence type experiments. However, it should be noted that employing such filtering reduces the overall success probability of the gate. In schemes where the gate is used to progressively construct resource states, this has the effect of incurring a polynomial overhead in resource requirements.

\begin{figure}[!htb]
\includegraphics[width=\columnwidth]{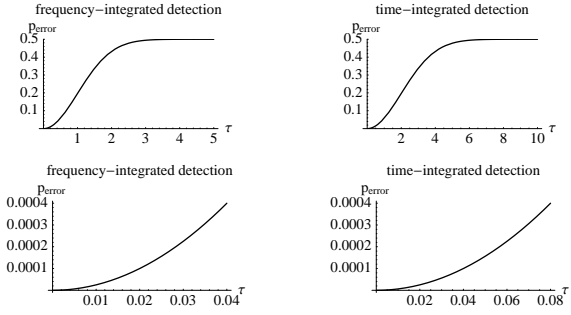}
\caption{Error probability ($p_\mathrm{error}$) against the degree of temporal mode-mismatch ($\tau$) in the limit of frequency-integrated detection (left) and time-integrated detection (right). The lower plots focus on the region of current fault-tolerant threshold estimates. $\tau$ is in units of photon bandwidth. In these limits the probability of detecting the \emph{success} signature is constant at 0.5.} \label{fig:int_det}
\end{figure}

In Fig.~\ref{fig:int_det} we consider the limits of \emph{frequency-integrated} and \emph{time-integrated} detection, where the photo-detectors' resolution and bandwidth in the respective domains are assumed to be infinite.

Since the parity gate operates non-determinisitcally, it is necessary to consider its behavior upon failure. Failure occurs when two photons are detected at one beamsplitter output port. Upon failure the $\ket{HV}$ and $\ket{VH}$ cases can be distinguished based on where the photons were detected. For example, if both photons are measured in the upper mode, the incident state must have been $\ket{VH}$. Thus, upon detecting both photons at one of the output ports, we have effectively performed a $Z$-measurement on both photons. The projection performed upon failure is unaffected by mode-mismatch, since we already have complete \emph{which-path} information, and mode-mismatch does not provide any additional distinguishing information.

The behavior of the parity gate can be subtly modified through the application of single qubit rotations prior to the gate. For example, in the type-II fusion gate described in Ref.~\cite{bib:BrowneRudolph05} both incident photons are first rotated by $45^\circ$. This has the effect of transforming the error model of Eqs.~\ref{eq:psi_ideal} and \ref{eq:psi_error} to
\begin{eqnarray}
\ket{\psi_\mathrm{ideal}}&=&\alpha_{HH}\ket{\phi_{HH}}+\alpha_{VV}\ket{\phi_{VV}}\nonumber\\
\ket{\psi_\mathrm{error}}&=&\alpha_{HV}\ket{\phi_{HV}}+\alpha_{VH}\ket{\phi_{VH}}
\end{eqnarray}
Thus, the error is no longer a phase error, but rather manifests itself as a probability of projecting into the wrong parity sub-space. Furthermore, upon failure the gate now performs an $X$-measurement (\emph{i.e.} in the $+/-$ basis) instead of a $Z$-measurement.
\begin{figure*}[!htb]
\includegraphics[width=0.8\textwidth]{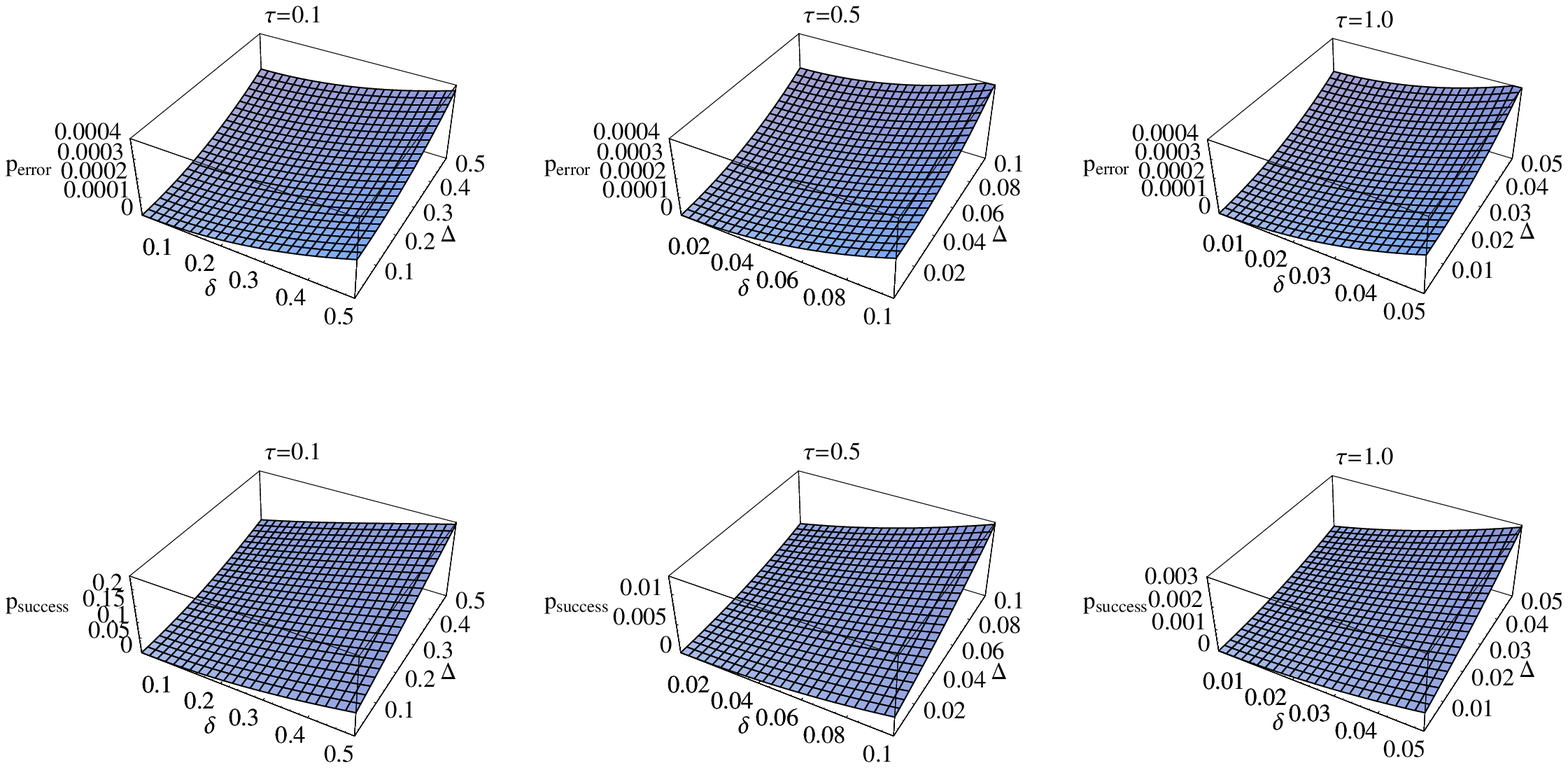}
\caption{Error probability ($p_\mathrm{error}$) and probability of detecting the \emph{success} signature ($p_\mathrm{success}$) against the photo-detectors' spectral resolution ($\delta$) and bandwidth ($\Delta$) for various degrees of temporal mode-mismatch ($\tau$). All quantities are in units of photon bandwidth.} \label{fig:p_delta_zoom}
\end{figure*}

\section{The cluster state model for quantum computation}
The standard model for quantum computation is very analogous to our understanding of classical circuits --  we prepare an input state, apply a series of gates, and measure the outputs. The cluster state \cite{bib:Raussendorf01,bib:Raussendorf03} model provides us with a completely different, yet computationally equivalent, model for quantum computing. We begin by preparing a maximally entangled, multi-qubit state, known as a \emph{cluster state}. Once a cluster state has been prepared, an arbitrary algorithm can be implemented by performing a sequence of single qubit measurements, which are trivial in an optical scenario. The order of these measurements and the choice of measurement bases determines the algorithm. Thus, cluster states act as a resource for universal quantum computation.

A cluster state can be represented as a graph. Nodes represent qubits initially prepared in the \mbox{$\ket{+}=(\ket{0}+\ket{1})/\sqrt{2}$} state. Vertices between nodes represent the application of controlled-sign ({\sc CZ}) gates between the respective qubits.

The cluster state model is particularly useful in the optical scenario, since it provides a means for performing LOQC far more efficiently \cite{bib:Nielsen04} than previous proposals. This is achieved by using non-deterministic {\sc CZ} gates to probabilistically produce a resource of small \emph{micro-clusters}. Larger clusters are constructed by progressively fusing micro-clusters onto the main cluster. When this fails the qubits being fused together are removed. When it succeeds we have successfully \emph{grown} the cluster. By ensuring the micro-clusters are sufficiently large we can always ensure that \emph{on average} the cluster grows as we repeat this process. Thus, we can grow arbitrarily large cluster states using physical resources which grow polynomially with the size of the final cluster.

\section{The redundantly-encoded cluster state scheme}
\begin{figure}[!htb]
\includegraphics[width=0.5\columnwidth]{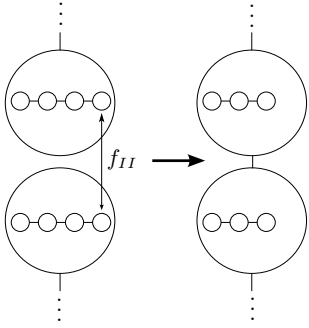}
\caption{The redundantly encoded cluster state scheme. Large circles represent logical cluster qubits, while smaller circles represent redundant physical qubits. A destructive {\sc CZ} gate is performed between two physical qubits of distinct logical qubits using an $f_{II}(H\otimes I)$ operation. This removes the physical qubits and introduces a {\sc CZ} operation between the logical qubits, thereby joining them in the cluster.} \label{fig:redundant_enc}
\end{figure}

We specifically consider the scheme described in Ref.~\cite{bib:BrowneRudolph05}, whereby each logical qubit is encoded using a redundant array of physical qubits. Specifically, $\ket{0}_L\equiv\ket{H}^{\otimes n}$ and $\ket{1}_L\equiv\ket{V}^{\otimes n}$, where $n$ is the level of encoding. We assume a resource of such states is available and therefore restrict ourselves to considering the errors introduced during the fusion processes. The {\sc CZ} gates are applied between one physical qubit from each logical qubits, referred to as the \emph{detachable} qubits. Because the physical qubits within each logical qubit are correlated, this is equivalent to performing the {\sc CZ} gate between the logical qubits. Desctructive {\sc CZ} gates can be implemented by applying a Hadamard gate to one qubit and applying a parity gate between them. Because we are utilizing redundant encoding, performing destructive {\sc CZ} gates does not destroy the logical qubits, but reduces their level of encoding by one, shown in Fig.~\ref{fig:redundant_enc}.

We now show this in detail. Consider a completely general state where we have factored out the logical qubits being fused,
\begin{eqnarray}
\ket{\psi}&=&\alpha_{HH}\ket{H}^{\otimes n}\ket{H}^{\otimes n}\ket{\phi_{HH}}\nonumber\\
&+&\alpha_{HV}\ket{H}^{\otimes n}\ket{V}^{\otimes n}\ket{\phi_{HV}}\nonumber\\
&+&\alpha_{VH}\ket{V}^{\otimes n}\ket{H}^{\otimes n}\ket{\phi_{VH}}\nonumber\\
&+&\alpha_{VV}\ket{V}^{\otimes n}\ket{V}^{\otimes n}\ket{\phi_{VV}}
\end{eqnarray}
Factorizing the detachable qubits and applying a Hadamard gate to the first qubit we obtain
\begin{eqnarray}
(H\otimes I)\ket{\psi}&=&\left[\alpha_{HH}\ket{H}^{\otimes n-1}\ket{H}^{\otimes n-1}\ket{\phi_{HH}}]\right\ket{H}\ket{H}\nonumber\\
&+&\left[\alpha_{HH}\ket{H}^{\otimes n-1}\ket{H}^{\otimes n-1}\ket{\phi_{HH}}]\right\ket{V}\ket{H}\nonumber\\
&+&\left[\alpha_{HV}\ket{H}^{\otimes n-1}\ket{V}^{\otimes n-1}\ket{\phi_{HV}}]\right\ket{H}\ket{V}\nonumber\\
&+&\left[\alpha_{HV}\ket{H}^{\otimes n-1}\ket{V}^{\otimes n-1}\ket{\phi_{HV}}]\right\ket{V}\ket{V}\nonumber\\
&+&\left[\alpha_{VH}\ket{V}^{\otimes n-1}\ket{H}^{\otimes n-1}\ket{\phi_{VH}}]\right\ket{H}\ket{H}\nonumber\\
&-&\left[\alpha_{VH}\ket{V}^{\otimes n-1}\ket{H}^{\otimes n-1}\ket{\phi_{VH}}]\right\ket{V}\ket{H}\nonumber\\
&+&\left[\alpha_{VV}\ket{V}^{\otimes n-1}\ket{V}^{\otimes n-1}\ket{\phi_{VV}}]\right\ket{H}\ket{V}\nonumber\\
&-&\left[\alpha_{VV}\ket{V}^{\otimes n-1}\ket{V}^{\otimes n-1}\ket{\phi_{VV}}]\right\ket{V}\ket{V}\nonumber\\
\end{eqnarray}
Following the parity gate we are left with
\begin{eqnarray} \label{eq:psi_f2_ideal}
f_{II}(H\otimes I)\ket{\psi}&=&\alpha_{HH}\ket{H}^{\otimes n-1}\ket{H}^{\otimes n-1}\ket{\phi_{HH}}\nonumber\\
&+&\alpha_{HV}\ket{H}^{\otimes n-1}\ket{V}^{\otimes n-1}\ket{\phi_{HV}}\nonumber\\
&+&\alpha_{VH}\ket{V}^{\otimes n-1}\ket{H}^{\otimes n-1}\ket{\phi_{VH}}\nonumber\\
&-&\alpha_{VV}\ket{V}^{\otimes n-1}\ket{V}^{\otimes n-1}\ket{\phi_{VV}}
\end{eqnarray}
which is equivalent to the application of a {\sc CZ} gate between the factored logical qubits.

\section{Error model for cluster states}
\begin{figure}[!htb]
\includegraphics[width=0.9\columnwidth]{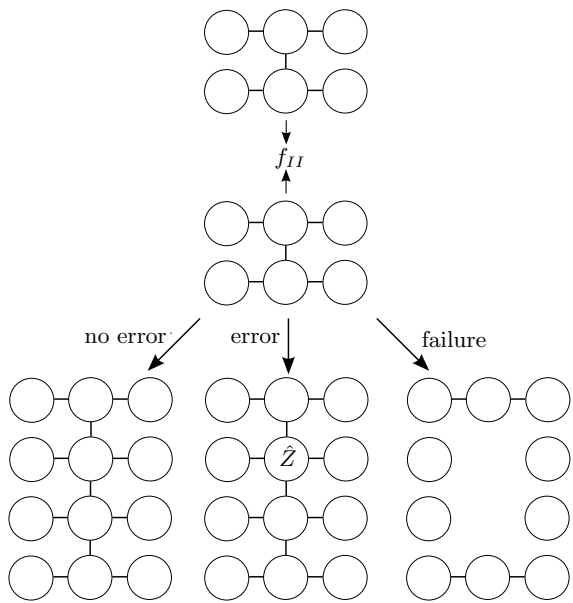}
\caption{Error model for the fusion gate in the construction of redundantly encoded cluster states. When the parity gate succeeds the two clusters are joined together. With some probability a $Z$-error will be introduced onto one of the the fused qubits. When the parity gate fails the fused qubits are removed from the clusters.} \label{fig:cluster_fig}
\end{figure}

We now consider the application of the parity gate error model to the redundantly encoded cluster state scheme. In the presence of mode-mismatch the gate has a probability of projecting into the wrong parity sub-space. Therefore,
\begin{eqnarray}
\ket{\psi_\mathrm{error}}&=&\left[\alpha_{HH}\ket{H}^{\otimes n-1}\ket{H}^{\otimes n-1}\ket{\phi_{HH}}\right]\nonumber\\
&+&\left[\alpha_{HV}\ket{H}^{\otimes n-1}\ket{V}^{\otimes n-1}\ket{\phi_{HV}}\right]\nonumber\\
&-&\left[\alpha_{VH}\ket{V}^{\otimes n-1}\ket{H}^{\otimes n-1}\ket{\phi_{VH}}\right]\nonumber\\
&+&\left[\alpha_{VV}\ket{V}^{\otimes n-1}\ket{V}^{\otimes n-1}\ket{\phi_{VV}}\right]
\end{eqnarray}
which differs from the ideal case (Eq.~\ref{eq:psi_f2_ideal}) through the application of a phase-flip to the first logical qubit. Thus, following fusion the state can be expressed
\begin{equation}
f_{II}(H\otimes I)\ket\psi=(1-p_\mathrm{error})\ket{C}\bra{C}+p_\mathrm{error}\hat{Z}_i\ket{C}\bra{C}\hat{Z}_i
\end{equation}
where $\ket{C}$ is the desired cluster state and $i$ denotes the fused qubit. This is simply a dephasing error model, as shown in Fig.~\ref{fig:cluster_fig}. It has been shown that quantum error correction is possible for such error models \cite{bib:NielsenDawson04}. Fault tolerant thresholds for a full Pauli error model with loss on cluster states have been estimated to be on the order of $10^{-4}$ \cite{bib:Dawson05}. Dephasing is a subset of this error model and can therefore be corrected for in principle. As before, achieving error probabilities within this threshold is possible, assuming sufficient control over detector characteristics and filtering. Once again, this comes at the expense of success probability, which incurs a polynomial physical resource overhead.

Upon gate failure, both physical qubits are effectively measured in the computational basis, which removes the respective logical qubits from the cluster, but does not destroy the remainder of the cluster. Because our model employs non-ideal detectors, which have finite bandwidth, it is also possible that less than a total of two photons are detected between the output ports of the parity gate. This is equivalent to photon loss. When this happens the affected logical qubits are irrecoverably destroyed. The remainder of the cluster can be recovered by measuring all neighboring qubits in the computational basis. In terms of the physical resource overhead, this is clearly more costly than a standard gate failure. However, the overhead is nonetheless polynomial, and scalable quantum computation is still possible in principle.

\section{Conclusion}
We constructed an error model for mode-mismatch in the parity gate, which forms the basis of several recent proposals for scalable linear optics quantum computing and other quantum optics experiments. This model was applied to the cluster state model for quantum computing. We related our results to current estimates for fault tolerant thresholds and found that mode-mismatch can be tolerated using existing quantum error correction techniques, assuming sufficient control over photo-detector characteristics and filtering. This comes at the expense of success probability, which affects the overall scaling of such schemes. However, the scaling of these schemes is polynomial with failure rate, and therefore in principle does not inhibit scalable linear optics quantum computing. While we specifically applied our model to a cluster state approach for LOQC, our model could easily be applied to other proposals where the parity gate is the fundamental building block.

\begin{acknowledgments}
We thank Daniel E. Browne, Terry Rudolph and Henry L. Haselgrove for helpful discussions. This work was supported by the Australian Research Council and the QLD State Government.
\end{acknowledgments}

\bibliography{paper}

\end{document}